\documentclass[english,aps,pra,reprint,noshowpacs,superscriptaddress]{revtex4-1}   
\usepackage[T1]{fontenc}	% should generally be included for better accented-word behavior
\usepackage[latin9]{inputenc}	% should generally be included for better accent behavior
\usepackage{geometry}		% for controlling page margins
\usepackage{graphicx}
\usepackage[above,below]{placeins}	% allows use of \FloatBar­rier command to force section barriers
\usepackage{times}
\usepackage{hyperref}  
\hypersetup{colorlinks=true,urlcolor=blue,citecolor=blue,linkcolor=blue}   
\usepackage{tabularx}
\usepackage{enumitem}

\begin{document}

\title{Initial impacts of the transformation of a large introductory lab course\\ focused on developing experimental skills and expert epistemology}
\author{H. J. Lewandowski}
\affiliation{Department of Physics, University of Colorado Boulder, Boulder, CO, 80309}
\affiliation{JILA, National Institute of Standards and Technology and University of Colorado Boulder, Boulder, CO, 80309}
\author{Daniel R. Bolton}
\affiliation{Department of Physics, University of Colorado Boulder, Boulder, CO, 80309}
\author{Benjamin Pollard}
\affiliation{Department of Physics, University of Colorado Boulder, Boulder, CO, 80309}
\affiliation{JILA, National Institute of Standards and Technology and University of Colorado Boulder, Boulder, CO, 80309}
\

\begin{abstract}
Recently, there has been increased attention to improving laboratory instruction at all levels.  At the introductory level, research results have shown differing levels of success based on the nature of the desired learning outcomes. In response to these findings, the University of Colorado's introductory physics lab course was transformed to improve students' development of experimental skills and experimental physics epistemology. We describe the details of the transformation process and initial self-reported learning gains from the first implementation of the transformed course.
\end{abstract}

\maketitle

\section{Introduction}

Laboratory courses can offer significant opportunities for engagement in the core scientific practices of experimental physics (e.g., asking questions, designing and carrying out experiments, analyzing data, developing and refining models, and presenting results to peers). The lab environment also has many affordances, which typically include extensive lab equipment, flexible classroom arrangements, low student/instructor ratios, and opportunities for collaborative work. Despite the abundant opportunities and resources in many laboratory courses, concerns are frequently raised about how effective such courses are at fulfilling their potential \cite{valueadded}. There are many calls to transform lab courses coming from the physics education community \cite{Holmes99,zwickl13}, the community of laboratory instructors \citep{aapt},  as well as national science policies promoting the retention of STEM majors and the development of the STEM workforce \cite{excel2012}. In particular, the American Association of Physics Teachers recently adopted the ``Recommendations for the Undergraduate Physics Laboratory Curriculum \citep{aapt}.'' These recommendations outline many desired learning outcomes for both the introductory lab and physics labs beyond the first year of college. 

Aligned with the AAPT guidelines to improve lab courses, we, at the University of Colorado Boulder (CU), have been working over the last several years on transforming all of our physics lab courses.  Recently, we began to focus on our large first-year lab (PHYS-1140), which serves over 700 engineering and physical science majors each semester. Prior to the transformation, this one-credit stand-alone course incorporated activities and apparatus originally developed in the 1960's, with only minor modifications in subsequent years. Apparatus included a simple pendulum and a parallel plate capacitor. Weekly activities were guided by detailed manuals and culminated in written reports. Faculty stated that the major learning goal of the course was teaching students how to propagate errors. This main learning goal, even if achieved, was not valued by either the students or the physics and engineering faculty members.  Therefore, we embarked on a process to completely transform the course. The transformation included defining new learning goals, modifying the structure and content of the course, and assessing student learning in several areas to monitor and improve the effectiveness of the course.

In the following sections, we (1) describe the process and outcomes of creating consensus learning goals for the transformed course, (2) outline the new course structure and activities, (3) present the methodology behind the two data sources for assessment, and (4) report initial outcomes of the first implementation of the transformed course along one dimension of learning: scientific communication. 
\vspace{-0.5cm}
\section{Consensus Learning Goals}

The transformation process started by establishing a new set of learning goals for the course. Because engineering majors make up a majority of students in the course, input was requested from faculty in the Physics Department and the College of Engineering and Applied Sciences (CEAS) at CU. All physics faculty members were invited to participate in an individual interview about the goals of the new course, and 10 participated. Data from the interviews were used to create a list of common learning goals.  Additionally, the Director the of Assessment and Accreditation in the CEAS distributed an online survey to all CEAS faculty asking for input on PHYS-1140. We received over 90 responses that were used to identify common themes. 

The goals desired by both the physics and CEAS faculty members were then discussed at three round-table meetings open to all faculty. The purpose was to build consensus for a tractable set of learning goals for PHYS-1140. Eighteen faculty members participated in at least one of these meetings. The outcome of this process was the following set of five broad learning goals and associated assessment instruments.

\begin{enumerate}[noitemsep, topsep=0pt]
	\item Students' epistemology of experimental physics should align with the expert view. \textit{Assessment: E-CLASS epistemology-focused items \cite{eclass}}
	\item Students should have a positive attitude about the course. \textit{Assessment: Course evaluations}
	\item Students should have a positive attitude about experimental physics. \textit{Assessment: E-CLASS affect-focused items}
	\item Students should be able to make a presentation quality graph showing a model and data. \textit{Assessment: Course artifacts}
	\item Students should demonstrate a set-like \cite{pmq} reasoning when evaluating measurements. \textit{Assessment: Physics Measurement Questionnaire\cite{pmq}}
\end{enumerate}

No goals included reinforcing physics concepts. Instead, faculty focused on goals that are unique to lab environments, such as scientific practices and views of experiential physics. These broad learning goals, as well as some additional structure goals (e.g., the new course should be sustainable even as many different faculty members rotate teaching the course), were used to guide decisions about both the structure and content of the transformed course. 
\vspace{-0.5cm}
\section{Transformed course components}

The structure and content of the transformed course are informed by the learning goals, structure goals, and institutional constraints of space, funding, and skills of the graduate teaching assistants (TAs). The course is structured around six 50-minute lectures and 12 two-hour lab activities. The 12 lab activities are grouped into four modules (skill building, mechanics, electronics, and optics). The first three labs are meant to develop capacity with basic experimental skills that are needed for the rest of the course. These include keeping an electronic lab notebook using a tablet and Microsoft OneNote, creating a graph with Excel embedded within OneNote, and being introduced to distributions of measurements and calculating uncertainties based on these distributions. During the nine subsequent labs, students work with additional equipment such as tennis ball launchers, capacitors, LEDs, and fiber optics. 

The students' grades are determined based on credit from clicker questions in lecture, online prelab activities, lab participation, lab notebooks, and completion of online assessments. Lab notebooks constitute 72\% of the final grade for the course. Each week, at the end of the lab session, the students create a PDF version of their lab notebook and upload it to the learning management system (Canvas). The notebooks are then graded online, where TAs make comments directly on the PDFs using detailed rubrics to assure consistency in grading among the 22 TAs. The use of lab notebooks represents a new component to the class, as previously students wrote only traditional lab reports using Mathematica and did not keep a lab notebook.

\section{Methodology}

To evaluate the transformed course's impact on students, we collected student artifacts (i.e., lab notebooks), administered surveys, and conducted focus group interviews. Surveys include the Colorado Learning Attitudes about Science Survey for Experimental Physics (E-CLASS) \citep{eclass} and the Physics Measurement Questionnaire (PMQ) \cite{pmq}. These are validated surveys that measure, respectively, students' views about experimental physics and their understanding of measurement uncertainty. Further, we probed students' affective responses to each of the lab activities and their self-reported learning gains in the course. To do so, we used both extra questions added to the post-instruction PMQ survey and focus group interviews. These latter two sources of data are the focus of this paper.

The questions added to the end of the PMQ probed various aspects of the course that were not directly assessed by E-CLASS and PMQ. These questions probed students' affective responses to each of the lab activities and self-reported learning gains on a few components of the course. The focus groups aimed to cover similar topics. The coupling of quantitative results on the extra PMQ questions, which all students answered, to the qualitative data extracted from a smaller subset of students in the focus groups allows us to capture all students' views of the course, as well as the possible underlying reasoning for those views. 
	
The post-instruction PMQ was administered during the last week of the course as a required online survey (1.5\% of their final grade). In total, 94\% of the students enrolled in the course competed the survey. To recruit participants for the focus groups, BP sent a recruitment email to all students in the course, and interested students were asked to provide their, availability, major, and (optionally) their gender.  BP formed five groups of five students based on respondents' gender, major, and scheduling constraints. Groups were intentionally homogeneous \cite{focus} in gender, and all but one was homogeneous with respect to major. A group of five women with either ``physics'' or ``other science'' majors was necessary due to the scheduling constraints of respondents.  The demographics of all study participants are shown in Table \ref{table:demo}.

The one-hour focus group sessions took place in the penultimate week of the semester, after students had completed all the lab activities in the course. Across all five focus groups, 18 students participated. BP conducted each session as a semi-structured interview, after having practiced the protocol once beforehand with a group of professional experimental physicists. Discussion topics ranged from comparing the course to other lab courses students had taken, reflecting on activities in the course that were the most and least enjoyable, and providing feedback on particular aspects of the course that corresponded to the transformation learning goals. BP took field notes during and after each session, and the sessions were audio and video recorded. For this study, BP selected transcript excerpts related to the goal of scientific communication.

We do point out there are several limitations to this study including the use of self-reported learning gains as an accurate measure of learning. We acknowledge that a more direct measure of student learning would be desirable, but argue, as others have \cite{Anaya1999}, that there is still value in self-reported gains, espcially in large classes with complex environments \cite{Douglass2012}. Additionally, we note that the focus group participants self-selected to participate in the activity and may not represent the student population as a whole. 
		
\begin{table}[ht]
\caption{Demographic information for study participants for the PMQ survey (N = 681) and focus groups (N = 18). Note that the Physics category includes both physics and engineering physics majors; the Other science
category includes (but is not limited to) biology, chemistry, and
math majors; and Nonscience includes both declared non-science
majors and students who are open option or undeclared.}
\centering % used for centering table
\begin{tabular}{l c c} % centered columns (4 columns)
\hline\hline %inserts double horizontal lines
Gender and Major & PMQ (\%) & Focus groups (\%)\\  % inserts table
%heading
\hline % inserts single horizontal line
Men & 73 & 61 \\ % inserting body of the table
Women & 26 & 39\\
Gender non-conforming & 1 & 0 \\
\hline
Physics & 12 & 22 \\
Engineering & 59 & 44 \\
Other science & 26 & 33 \\
Non-science & 3 & 0\\
[1ex] % [1ex] adds vertical space
\hline %inserts single line
\end{tabular}
\label{table:demo} % is used to refer this table in the text
\end{table}

\begin{figure}
\begin{center}

	\begin{minipage}{.5\textwidth}
		\includegraphics[width=\linewidth]{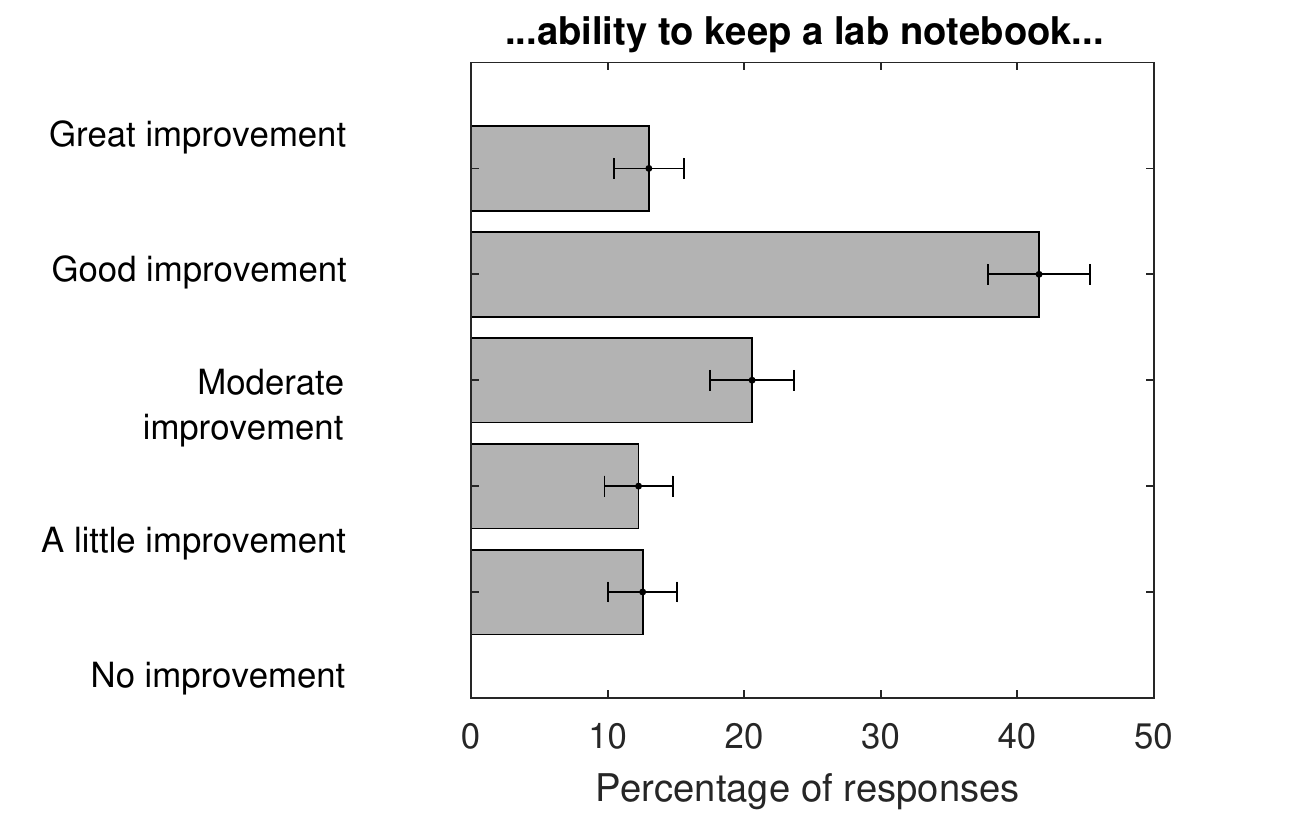}
	\end{minipage}
	\begin{minipage}{.5\textwidth}
	\includegraphics[width=\linewidth]{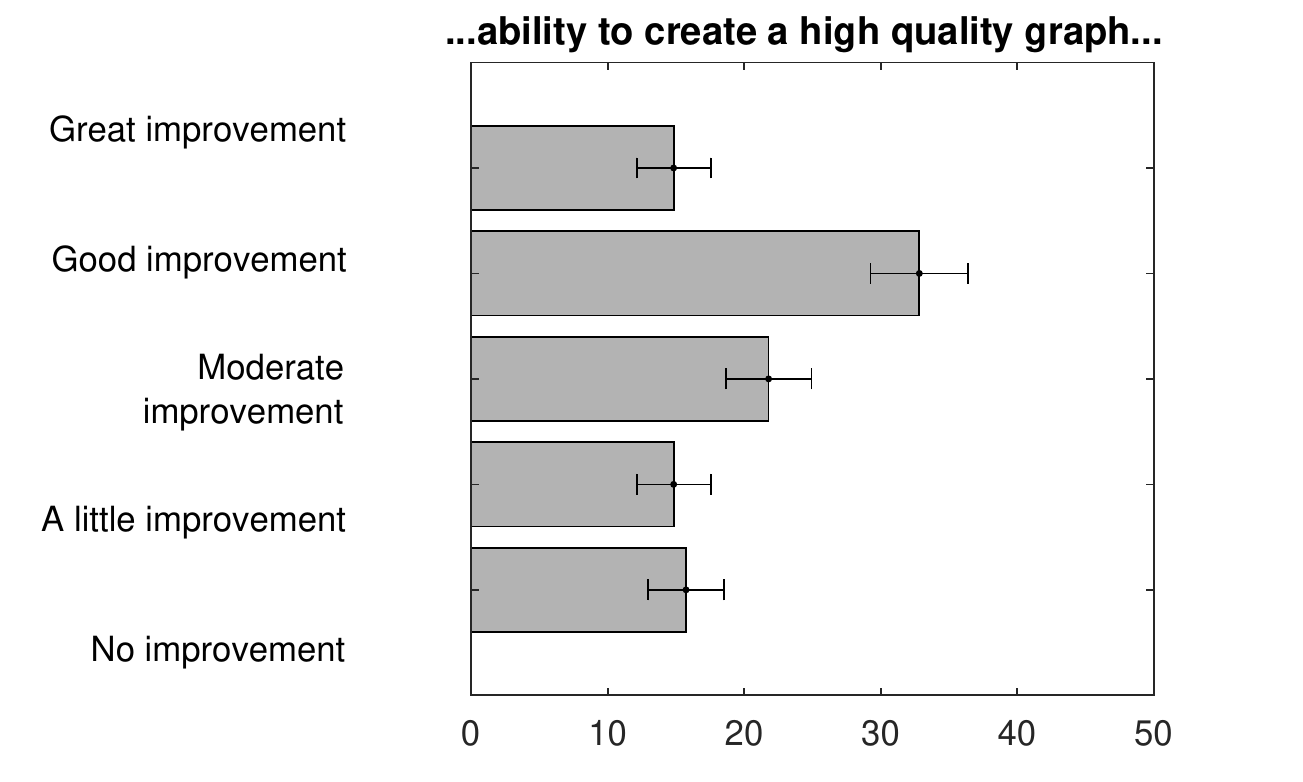}
		\end{minipage}
		\label{fig:graph1}
		\end{center}
		\caption{Student responses to Q1 (top) and Q2 (bottom). Error bars represent the 95\% confidence interval.   }
	
\end{figure}

\begin{figure}
\begin{center}
	\begin{minipage}{.5\textwidth}
		\includegraphics[width=\linewidth]{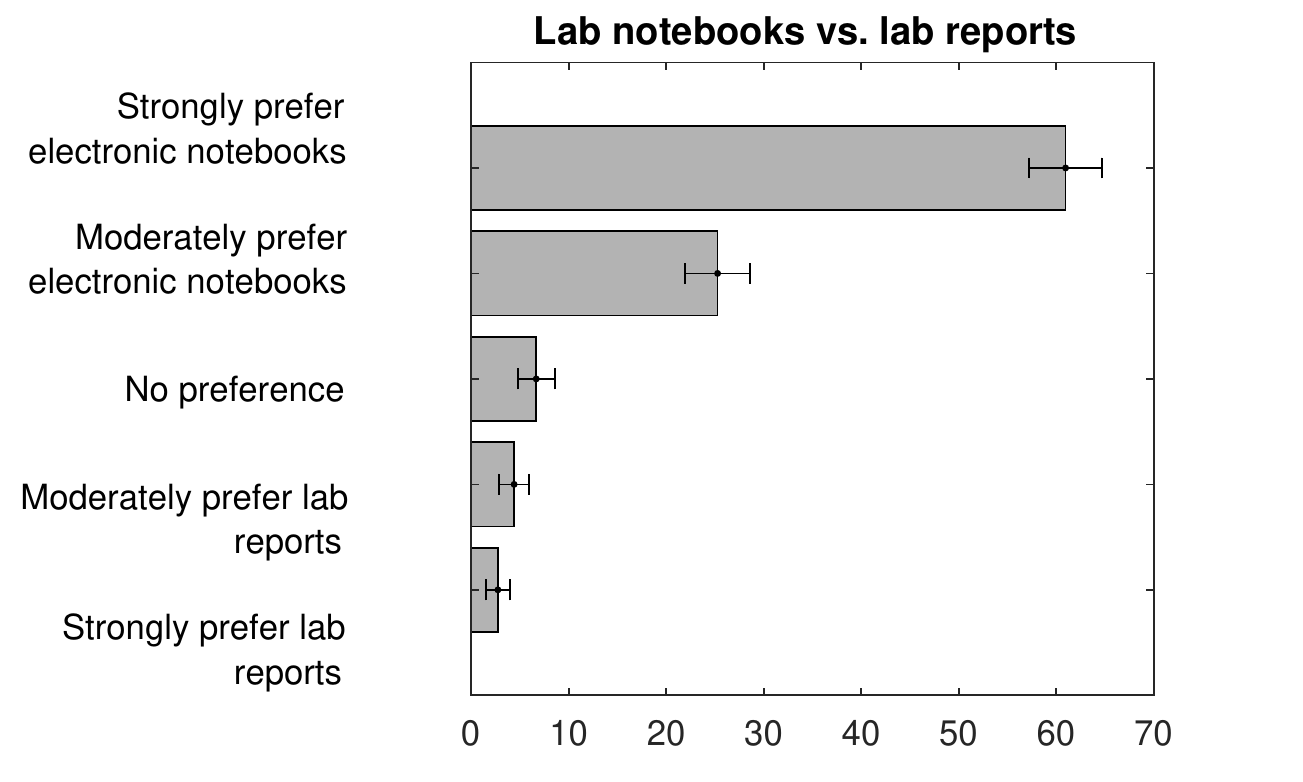}
	
	\end{minipage}
	\begin{minipage}{.5\textwidth}
	\includegraphics[width=\linewidth]{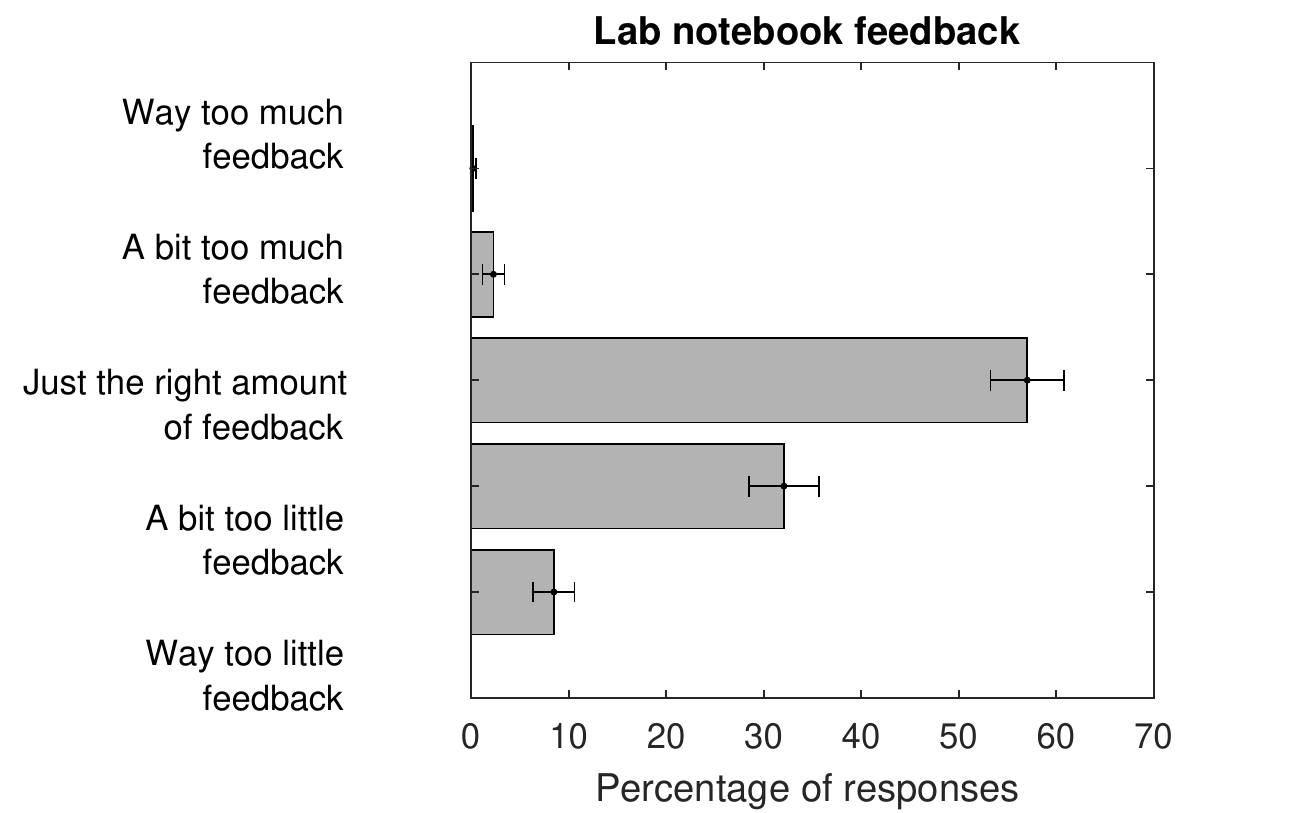}
		\end{minipage}
		\label{fig:graph2}
		\end{center}
		\caption{Student responses to Q3 (top) and Q4 (bottom). Error bars represent the 95\% confidence interval.  }
	
\end{figure}

\section{Results and discussion}
For the study described here, we limit ourselves to the topic of scientific communication, specifically lab notebook use and creation of graphs in the notebooks. There were four questions on the post-survey dealing with lab notebooks. 

\begin{enumerate}[noitemsep,topsep=0pt]
\item[Q1] How much did your ability to keep a lab notebook improve as a result of taking PHYS 1140?
\item[Q2] How much did your ability to create a high quality graph improve as a result of taking PHYS 1140?
\item[Q3] In PHYS-1140, you turned in electronic notebooks instead of lab reports. To what extent do you prefer one or the other?
\item[Q4] How do you feel about the amount of feedback you received on the electronic notebooks you turned in?
\end{enumerate}
These Likert-style questions were designed to probe students' perceptions of their own learning, preferences for notebooks or reports, and perceived adequacy of feedback.

As can be seen in Figs. 1 and 2, over 70\% of students reported moderate-to-great improvement in their capacity to create a high quality graph, or to maintain a lab notebook. Additionally, over 80\% said they preferred lab notebooks to reports. 

To better understand why students had favorable views about lab notebooks, we draw upon common themes from the focus groups. One aspect of student reasoning behind their responses was a connection to professional practice or a future job.
\vspace{-0.3cm}
\begin{itemize}
\item[] \emph{I imagine that [electronic lab notebook use] is more realistic to professional experience as well. I'm sure they use embedded Excel plots and linear fit graphs and all that. You're not going to draw that into a [non-electronic] lab notebook.}
\end{itemize}

\begin{itemize}
\item[]\emph{This was the first class I've had where you used technology like that, and I feel like that is pretty ... when you get a job and move on, it's gonna be more and more computer based compared to writing old-school notes.}
\end{itemize}

Another common theme was that students preferred lab notebooks because, compared to reports, they were easier to use:
\vspace{-0.3cm}
\begin{itemize}

\item[]\emph{[Electronic notebooks] were definitely quicker...once you got settled in to using the surface tablet, after lab 1,2,3, you pretty much got the hang of it. So you started to understand more about how to use it and it seemed like it was, like I said, more convenient. Less time consuming. }
\end{itemize}

Finally, along the lines of improving notebook keeping and graphing, students often discussed improved proficiency with OneNote and Excel software. 

\begin{itemize}
\item[]\emph{Going into [PHYS-1140], I was unsure. I didn't really know how to use any of the software. ... I was a little worried about that. Also Excel. I'm not super experienced with [it]. So the first few labs, there was definitely a bit of learning that I had to do.}
\end{itemize} 
\begin{itemize}
\item[]\emph{...my lab partner wasn't good at Excel so I saw her throughout the semester get way better at it. She became more familiar with adding graphs, and all that other stuff.}
\end{itemize}

Despite multiple indicators that students had a positive experience with the notebooks, there was one obvious area for improvement: the amount of feedback the students received from the TAs. While most students (57\%) felt that they received the right amount of feedback on their notebooks, about 40\% said they received a too little feedback. Based on these responses, we plan to modify the grading rubrics to allow TAs to more easily give students additional feedback.
\vspace{-0.5cm}
\section{Conclusions and future research}

In response to national calls \citep{aapt} to reform laboratory education and a local desire to improve student learning in the first-year physics lab course at CU, we developed a new set learning goals and used these to guide a complete transformation of our PHYS-1140 course. Initial results from survey questions and focus groups indicate that introductory labs can engage students in the scientific practice of keeping a lab notebook, and that students both preferred this mode of communication and believed they improved their ability to keep an electronic notebook as a result of the course. The results also demonstrate that additional guidance for the TAs is necessary to increase the amount of feedback they provide to students on their notebooks. 
	
In the future, these findings will be correlated with responses to E-CLASS items related to students' views about the role of communication in physics, as well as qualitative analysis of student notebooks to understand their notebook competency beyond the software. By looking across multiple sources of data, we can get a more comprehensive picture of students' learning about scientific communication via lab notebooks.
	
\vspace{-0.9cm}
\acknowledgments{We thank the entire transformation team including M. Dubson, A. Ellzey, R. Hobbs, M. Schefferstein, C. West, and D. Woody for their valuable contributions and the physics and engineering faculty for their input on learning goals. Additionally, we thank the focus group participants for their insights into the course. This work is supported by the NSF under grant PHYS-1734006, the TRESTLE program DUE-1525331, the office of the Assoc. Dean for Education of the CEAS, and the College of Arts and Sciences at the University of Colorado.}

\vspace{-0.5cm}
\bibliographystyle{apsrev}  	% supercedes the longbibliography option, so leave commented out if you want to display article titles
\bibliography{PERC2018}  	% don't include the .bib suffix

\end{document}